# A new exact algorithm for solving single machine scheduling problems with learning effects and deteriorating jobs

Soheyl Khalilpourazari[1]

**Department of Industrial Engineering, Faculty of Engineering, Kharazmi University, Tehran, Iran**
**Phone: +98 (21) 88830891, Fax: +98 (21) 88329213, e-mail: std_khalilpourazari@khu.ac.ir**

Mohammad Mohammadi

**Department of Industrial Engineering, Faculty of Engineering, Kharazmi University, Tehran, Iran**
**Phone: +98 (21) 88830891, Fax: +98 (21) 88329213, e-mail: mohammadi@khu.ac.ir**

**Abstract:**

In this paper, the single machine scheduling problem with deteriorating jobs and learning effects are considered, which is shown in the previous research that the SDR method no longer provides an optimal solution for the problem. In order to solve the problem, a new exact algorithm is proposed. Various test problems are solved to evaluate the performance of the proposed heuristic algorithm using different measures. The results indicate that the algorithm can solve various test problems with small, medium and large sizes in a few seconds with an error around 1% where solving the test problems with more than 15 jobs is almost impossible by examining all possible permutations in both complexity and time aspects.

---

[1] Corresponding author

## 1. Introduction and literature review

In classical scheduling problems, it is assumed that the processing time of jobs is constant and known. In many real-world conditions, the processing time of jobs may influence by many factors. Deterioration and learning effect are important factors which can influence the processing time of jobs. By the deterioration, we mean that the processing time of jobs is a function of their starting times. By learning effect, we mean that the processing time of a job depends on its position in the sequence. In the last decade many researchers paid attention to scheduling problems with deteriorating jobs and learning effect. Wen-Chiung Lee (2004) and CHENG Ming-bao, SUN Shi-jie (2006) considered deterioration and learning effect simultaneously and proved that the makespan problem remains polynomial solvable. To the best of our knowledge papers which considered deteriorating jobs or learning effect or considered both deteriorating jobs and learning effect simultaneously include Lee and Wu (2008, 2009), Cheng et al. (2008), Mosheiov (2008), Toksar and Guner (2008), Tang and Liu (2008, 2009), Lee et al. (2009), Wang (2009), Wang et al. (2009), Wang and Liu(2009), Wang and Guo (2010), Wang et al. (2010), Wang and Wang (2011,2012), Sun et al. (2012) and more recent papers include Low and Lin (2013), Wang and Wang (2013), Xu et al. (2014), Wang and Wang (2014), Yin et al. (2014), and Mohammadi and Khalilpourazari (2017).

Wen-Chiung Lee (2004) considered deterioration and learning effect simultaneously in single machine scheduling problem. He proved that the minimization problems of makespan and total follow time remain polynomial solvable. He assumed jobs processing time as $p_{i,r} = \alpha_i t r^a$ where $p_0$ is basic processing time of job and $\alpha_i$ is the deterioration rate of job $i$ which is scheduled in $r^{th}$ position and $a \leq 0$ is learning index. He proved that for the problem $1 \,|p_{j,r} = (p_0 + \alpha_i t)r^a|\, C_{max}$ the well-known SDR method no longer provides the optimal solution. Ming-bao and Shi-jie (2006) considered deterioration and learning effect simultaneously in single machine scheduling problem where the processing time of jobs followed as $p_{j,r} = (a_j + b_j t)\alpha^{r-1}$ they proposed polynomial solutions for makespan, maximum lateness minimization, total flow times minimization problems under special cases. For the third special case they proved by an example that for the problem $1 \,|p_{j,r} = (a_0 + b_j t)\alpha^{r-1}|\, C_{max}$ the classical methods no longer provide the optimal solution.

In this paper, we proposed a new algorithm to solve the problem $1|p_{j,r} = (a_0 + b_j t)\alpha^{r-1}|C_{max}$ which Ming-bao and Shi-jie (2006) proved SDR method no longer provides optimal solution. We first present sent the algorithm's solution procedure and then we solve various test problems. We obtain the optimal solution for each test problems examining all possible permutations of jobs then we validate the solutions obtained using two proposed algorithms. The results indicate that in most cases the algorithm obtains optimal solution and in other test problems the error percentage for the algorithm is around 1%. We use two measures (error percentage, Cpu-Time) to evaluate algorithm's performance in obtaining optimal or near optimal solutions and the time needed to solve the test problems.

## 2. Problem Definition and formulation

In this paper, we consider a model which consider the deterioration rate and learning effect simultaneously. In the model, there are $n$ jobs ready to be processed on a single machine. The assumptions are as follows:

1- The jobs are available and ready for processing at $T > 0$.

2- The machine can handle only one job at a special time $T > 0$.

3- Preemption is not allowed.

4- The jobs are independent.

**The first model:**

We consider the model proposed by Cheng Ming-bao and Sun Shi-jie(2006), which denotes $p_{j,r}$ as the processing time of job $J_j$ when it is scheduled in $r^{th}$ position in a sequence with normal processing time $a_0$. So we have:

$$p_{j,r} = (a_0 + b_j t)\alpha^{r-1} \tag{1}$$

Where $b_j$ is detethe rioration rate of job $J_j$ and $0 < a \le 1$ is the learning index and $t$ is the starting time of processing on job $J_i$.

**Theorem1:** for the problem $1|p_{j,r} = (a_0 + b_j t)\alpha^{r-1}|C_{max}$ and stating time $T \geq 0$, the optimal or a near optimal sequence can be obtained by following 3 steps named A, B and C:

**A) Finding the first sequence:**

**Step1-a:**

If $a_0 \geq \frac{T(\alpha^{r-1}-\alpha^r)}{\alpha^{2r-1}}$ then put the job with the **largest** deterioration rate at $r^{th}$ position. (LDR)

Or

If $a_0 < \frac{T(\alpha^{r-1}-\alpha^r)}{\alpha^{2r-1}}$ then put the job with the **smallest** deterioration rate at $r^{th}$ position. (SDR)

**Step2-a:**

Then calculate the new T (completion time of the last job) and check the two conditions mentioned above and put the second job after the first job and go to step1-a. Do this to remaining jobs to put them in sequence. Name this sequence as sequence1. (Note: for the first job we have $r = 1$.)

**B) Finding the second sequence:**

**Step1-b: (just for the first job)**

If $a_0 \geq \frac{T(\alpha^{r-1}-\alpha^r)}{\alpha^{2r-1}}$ then put the job with the **second largest** deterioration rate at $1_{st}$ position. (LDR)

Or

If $a_0 < \frac{T(\alpha^{r-1}-\alpha^r)}{\alpha^{2r-1}}$ then put the job with the **smallest** deterioration rate at $1_{st}$ position. (SDR)

Calculate T (completion time of last job (the first job)) and then:

**Step2-b:**

If $a_0 \geq \frac{T(\alpha^{r-1}-\alpha^r)}{\alpha^{2r-1}}$ then put the job with the **largest** deterioration rate at $r^{th}$ position. (LDR)

Or

If $a_0 < \frac{T(\alpha^{r-1}-\alpha^r)}{\alpha^{2r-1}}$ then put the job with the **smallest** deterioration rate at $r^{th}$ position. (SDR)

**Step3-b:**

Then calculate the new T (completion time of the last job) and check the two conditions mentioned in step2-b and put third job after the second job. Do this to remaining jobs to put them in sequence using step2-b. Name this sequence as sequence2. (Note: for first job we have $r = 1$.)

**C) The best sequence**

Calculate the Makespan for sequence1 and sequence2, Then the sequence with lower Makespan is the best near-optimal sequence.

**Proof:** suppose two schedules of jobs $\rho = [S_1, J_k, J_l, S_2]$ and $\rho' = [S_1, J_l, J_k, S_2]$ where $S_1$, $S_2$ are partial sequences. We assume that there are $r - 1$ job in $S_1$ sequence. by considering $\rho$ then $J_k, J_l$ are the $r^{th}$ and the $(r+1)^{th}$ job in sequence. Let $T$ to denote the completion time of the last job in $S_1$.

Under $\rho$ sequence we have:

$$C_k(\rho) = T + (a_0 + b_k t)\alpha^r = a_0 \alpha^{r-1} + T(1 + b_k \alpha^{r-1}) \tag{2}$$

$$C_l(\rho) = C_k(\pi) + (a_0 + b_l C_k(\pi))\alpha^r = a\alpha^r + [a_0\alpha^{r-1} + T(1+b_k\alpha^{r-1})](1+b_l\alpha^r) \tag{3}$$

Similarly under $\rho'$ we have:

$$C_l(\rho') = a_0 \alpha^{r-1} + T(1 + b_l \alpha^{r-1}) \tag{4}$$

$$C_k(\rho') = a_0\alpha^r + [a_0\alpha^{r-1} + T(1+b_l\alpha^{r-1})](1+b_k\alpha^r) \tag{5}$$

To show that $\rho'$ dominates $\rho$ we have:

$C_k(\rho') - C_l(\rho)$
$$= \{a_0\alpha^r + a_0\alpha^{r-1} + a_0\alpha^{r-1}b_k\alpha^r + T + Tb_k\alpha^r + Tb_l\alpha^{r-1} + Tb_l\alpha^{r-1}b_k\alpha^r\}$$
$$- \{a_0\alpha^r + a_0\alpha^{r-1} + a_0\alpha^{r-1}b_l\alpha^r + T + Tb_l\alpha^r + Tb_k\alpha^{r-1} + Tb_l\alpha^{r-1}b_k\alpha^r\} \geq 0$$

so:

$$(b_k - b_l)(T\alpha^r - T\alpha^{r-1} + a_0\alpha^{2r-1}) \geq 0$$

We have to consider two conditions:

1) If $(T\alpha^r - T\alpha^{r-1} + a_0\alpha^{2r-1}) \geq 0$ then we have:

$$a_0 \geq \frac{T(\alpha^{r-1} - \alpha^r)}{\alpha^{2r-1}} \Rightarrow a_0 \geq \frac{T(1-\alpha)}{\alpha^r}$$

2) If $(T\alpha^r - T\alpha^{r-1} + a\alpha^{2r-1}) < 0$ then we have:

$$a_0 < \frac{T(\alpha^{r-1} - \alpha^r)}{\alpha^{2r-1}} \Rightarrow a_0 < \frac{T(1-\alpha)}{\alpha^r}$$

In the two following conditions, sequence $\rho$ dominates $\rho'$:

1- If $a_0 \geq \frac{T(\alpha^{r-1}-\alpha^r)}{\alpha^{2r-1}}$ it implies that $(T\alpha^r - T\alpha^{r-1} + a_0\alpha^{2r-1}) \geq 0$ then we have $(b_k - b_l) \geq 0$ to complete the proof. From $(b_k - b_l) \geq 0$ we have $b_k \geq b_l$ and by considering $\rho$ sequence, it is clear that the sequence which we put the job with the largest deterioration rate at the first, $(\rho)$, is better than the sequence $(\rho')$ which we have put the job with the smallest deterioration rate at the first, therefore in this condition $\left(a_0 \geq \frac{T(\alpha^{r-1}-\alpha^r)}{\alpha^{2r-1}}\right)$ we have to put the job with the largest deterioration rate at the first (when we have two jobs), The Largest Deterioration Rate (LDR) method.

2- If $a_0 < \frac{T(\alpha^{r-1}-\alpha^r)}{\alpha^{2r-1}}$ then $(T\alpha^r - T\alpha^{r-1} + a_0\alpha^{2r-1}) < 0$ so we have $(b_k - b_l) < 0$ to complete the proof. Therefore we have $b_k < b_l$ and by considering $\rho$ sequence, it is clear that the sequence $(\rho)$ which we have put the job with the smallest deterioration rate at the first, is better than sequence $(\rho')$ which we put the job with the largest deterioration rate at the first, therefore in this condition $\left(a_0 < \frac{T(\alpha^{r-1}-\alpha^r)}{\alpha^{2r-1}}\right)$ we have to put the job with the smallest deterioration rate at the first (when we have two jobs), The well-known Smallest Deterioration Rate (SDR) method. This completes the proof.

We proposed a new heuristic algorithm to solve the problem:

## 3. Heuristic Algorithm:

### A) Finding the first sequence:

**Step1-a:**

If $a_0 \geq \frac{T(\alpha^{r-1}-\alpha^r)}{\alpha^{2r-1}}$ then put the job with the largest deterioration rate at $r^{th}$ position. (LDR method)

Or

If $a_0 < \frac{T(\alpha^{r-1}-\alpha^r)}{\alpha^{2r-1}}$ then put the job with the smallest deterioration rate at $r^{th}$ position. (SDR method)

**Step2-a:**

Then calculate the new T (completion time of the last job) and check the 2 conditions mentioned above and put the second job after first job and go to step1-a. Do this to remaining to put them jobs in sequence. Name this sequence as sequence1. (Note: for the first job we have $r = 1$.)

### B) Finding the second sequence:

**Step1-b: (just for first job)**

If $a_0 \geq \frac{T(\alpha^{r-1}-\alpha^r)}{\alpha^{2r-1}}$ then put the job with the <u>second</u> largest deterioration rate at $1_{st}$ position. (LDR)

Or

If $a_0 < \frac{T(\alpha^{r-1}-\alpha^r)}{\alpha^{2r-1}}$ then put the job with the smallest deterioration rate at $1_{st}$ position. (SDR)

Calculate T (completion time of last job (the first job)) and then:

**Step2-b:**

If $a_0 \geq \frac{T(\alpha^{r-1}-\alpha^r)}{\alpha^{2r-1}}$ then put the job with the largest deterioration rate at $r^{th}$ position. (LDR)

Or

If $a_0 < \frac{T(\alpha^{r-1}-\alpha^r)}{\alpha^{2r-1}}$ then put the job with the smallest deterioration rate at $r^{th}$ position. (SDR)

**Step3-b:**

Then calculate the new T (completion time of the last job) and check the two-condition mentioned in step2-b and put the third job after the second job. Do this to remaining jobs to put them in sequence using step2-b. Name this sequence as sequence2. (Note: for first job we have $r = 1$.)

**C) The best sequence**

Calculate the Makespan for sequence1 and sequence2 and then the sequence with minimum Makespan is the best near optimal or optimal sequence.

**4. Performance evaluation**

We have solved various test problems with different sizes to compare solutions obtained from three solution methods for the problem $1 \,|p_{j,r} = (a_0 + b_j t)\alpha^{r-1}|\, C_{max}$:

We consider three solution methods as follows:

1- Calculating Makespan for all possible permutation of jobs and find the best sequence with the minimum Makespan.

2- The well-known Smallest Deterioration Rate (SDR) method.

3- Our proposed heuristic algorithm.

We obtained the best sequence of jobs for each test problem by calculating Makespan of all possible permutations and compared the results with our proposed algorithm and SDR method to find out which solution method is better, our proposed heuristic algorithm and the SDR method. In order to evaluate the performance of the solution methods the error percentage and Cpu-time measures are considered to compare all three solution methods, for more information see (Khalilpourazari and Pasandideh 2018; Khalilpourazari et al. 2018; Fazli-Khalaf et al. 2017; Khalilpourazari and Khalilpourazary 2018a; Khalilpourazari and Khalilpourazary 2018b; Khalilpourazari and Mohammadi 2016; Khalilpourazay et al. 2014; Khalilpourazari and

Khalilpourazary 2017; Pasandideh and Khalilpourazari 2018). We used MATLAB software to solve the test problems with three solution methods using a laptop with i7 Cpu and 8 GB of RAM.

Table 1 presents the distribution of parameters and the deterioration rate of jobs for each test problem presented in Table 2, the best sequence obtained by each solution method presented in Table 3 and the solution detailed results are presented in Table (4):

Table 1: The distribution of parameters

| parameter | distribution |
|---|---|
| $b_j$ | ~uniform(0,6) |
| $a_0$ | ~uniform(0.5,2.5) |
| $t$ | ~uniform(0.5,1.5) |
| $\alpha$ | ~uniform(0,1) |

Table 2: Deterioration rate of jobs for each test problem

| Number of jobs | Jobs deterioration rate |
|---|---|
| n=2 | $b_1 = 4.82, b_2 = 2.98$ |
| n=3 | $b_1 = 5.58, b_2 = 4.10, b_3 = 1,90$ |
| n=4 | $b_1 = 0.59, b_2 = 1.02, b_3 = 2.42, b_4 = 0.31$ |
| n=5 | $b_1 = 0.59, b_2 = 0.47, b_3 = 1.69, b_4 = 3.69, b_5 = 3.63$ |
| n=6 | $b_1 = 0.40, b_2 = 3.81, b_3 = 4.71, b_4 = 2.96, b_5 = 0.53, b_6 = 1.32$ |
| n=7 | $b_1 = 5.15, b_2 = 3.28, b_3 = 5.44, b_4 = 1.52, b_5 = 4.45, b_6 = 1.87, b_7 = 4.51$ |
| n=8 | $b_1 = 4.69, b_2 = 0.06, b_3 = 3.25, b_4 = 2.22, b_5 = 5.66, b_6 = 0.84, b_7 = 2.34, b_8 = 2.78$ |
| n=9 | $b_1 = 0.46, b_2 = 2.77, b_3 = 3.32, b_4 = 1.78, b_5 = 4.47, b_6 = 1.03, b_7 = 5.17, b_8 = 1.72, b_9 = 4.47$ |
| n=10 | $b_1 = 3.18, b_2 = 5.28, b_3 = 1.09, b_4 = 4.57, b_5 = 1.68, b_6 = 0.54, b_7 = 3.42, b_8 = 2.64, b_9 = 2.64, b_{10} = 3.67$ |

Table 3: The best sequence obtained by each solution method

| Number of jobs | Solution Method | Best Sequence |
|---|---|---|
| n=2 | Full numerate | $J_1, J_2$ |
| | SDR method | $J_2, J_1$ |
| | Our proposed algorithm | $J_1, J_2$ |
| n=3 | Full numerate | $J_1, J_3, J_2$ |
| | SDR method | $J_3, J_2, J_1$ |
| | Our proposed algorithm | $J_1, J_3, J_2$ |
| n=4 | Full numerate | $J_3, J_1, J_4, J_2$ |
| | SDR method | $J_4, J_1, J_2, J_3$ |
| | Our proposed algorithm | $J_3, J_2, J_4, J_1$ |
| n=5 | Full numerate | $J_4, J_2, J_1, J_3, J_5$ |
| | SDR method | $J_2, J_1, J_3, J_5, J_4$ |
| | Our proposed algorithm | $J_4, J_2, J_1, J_3, J_5$ |
| n=6 | Full numerate | $J_3, J_1, J_5, J_6, J_4, J_2$ |
| | SDR method | $J_1, J_5, J_6, J_4, J_2, J_3$ |
| | Our proposed algorithm | $J_3, J_1, J_5, J_6, J_4, J_2$ |
| n=7 | Full numerate | $J_1, J_4, J_6, J_2, J_5, J_7, J_3$ |
| | SDR method | $J_4, J_6, J_2, J_5, J_7, J_1, J_3$ |
| | Our proposed algorithm | $J_1, J_4, J_6, J_2, J_5, J_7, J_3$ |
| n=8 | Full numerate | $J_5, J_2, J_6, J_4, J_7, J_8, J_3, J_1$ |
| | SDR method | $J_2, J_6, J_4, J_7, J_8, J_3, J_1, J_5$ |
| | Our proposed algorithm | $J_5, J_2, J_6, J_4, J_7, J_8, J_3, J_1$ |
| n=9 | Full numerate | $J_5, J_1, J_6, J_8, J_4, J_2, J_3, J_9, J_7$ |
| | SDR method | $J_1, J_6, J_8, J_4, J_2, J_3, J_5, J_9, J_7$ |
| | Our proposed algorithm | $J_9, J_1, J_6, J_8, J_4, J_2, J_3, J_5, J_7$ |
| | Full numerate | $J_7, J_6, J_3, J_5, J_8, J_9, J_1, J_{10}, J_4, J_2$ |

| n=10 | SDR method | $J_6, J_3, J_5, J_8, J_9, J_1, J_7, J_{10}, J_4, J_2$ |
|---|---|---|
| | Our proposed algorithm | $J_4, J_6, J_3, J_5, J_8, J_9, J_1, J_7, J_{10}, J_2$ |

Table 4: Performance evaluation of the algorithms

| Number of jobs | Makespan | | | Cpu-time (s) | | | Error percentage | |
|---|---|---|---|---|---|---|---|---|
| | Full numerate | Our proposed algorithm | SDR | Full numerate | Our proposed algorithm | SDR | Our proposed algorithm | SDR |
| n=2 | 24.975 | 24.975 | 26.265 | 0 | 0 | 0 | 0 | 5.16 |
| n=3 | 82.948 | 82.9485 | 93.334 | 0 | 0 | 0 | 0 | 12.52 |
| n=4 | 16.780 | 16.864 | 21.159 | 0 | 0 | 0 | 0.50 | 26.09 |
| n=5 | 81.458 | 81.458 | 105.765 | 0 | 0 | 0 | 0 | 29.84 |
| n=6 | 185.756 | 185.756 | 249.031 | 0.1 | 0 | 0 | 0 | 34.06 |
| n=7 | 3836.419 | 3836.419 | 4361.866 | 1.01 | 0 | 0 | 0 | 13.69 |
| n=8 | 1020.844 | 1020.844 | 1435.403 | 7.97 | 0 | 0 | 0 | 40.60 |
| n=9 | 2162.115 | 2162.115 | 2691.113 | 101.87 | 0 | 0 | 0 | 24.46 |
| n=10 | 4277.653 | 4323.567 | 5219.636 | 917.10 | 0 | 0 | 1.70 | 22.02 |

As in Table 4, the mean error percentage of the proposed algorithm is 0.24 percent, while the mean error percentage for SDR method is 23.16 percent which shows that the proposed algorithm performs better than SDR method. On the other hand from comparing algorithms Cpu-Times it takes too long to calculate the Makespan for all possible permutations and find the best sequence, for example for $n = 10$ jobs we have to calculate the Makespan for $10! = 3628800$ sequences which takes 917.10 seconds, note that our computer with an i7 Cpu and 8GB of RAM can calculate the Makespan for 4817 sequences in just one second, considering this to find the best sequence for $n = 15$ jobs we have to calculate the Makespan for $15! = 1307674368000$ sequences which means it takes almost 9 years to find the best solution by examining all possible permutations, but the proposed algorithm can find a very near optimal sequence with a tight approximation and an error around 1% which means that the proposed algorithm is efficient in both Cpu-Time and error percentage comparing to other solution methods.

## 5. Conclusion and future research:

In this paper, a new heuristic algorithm is proposed to solve single machine scheduling problem with time-dependent processing times and learning effects. We evaluate the performance of the heuristic algorithm by solving various test problems using different measures. We show that the proposed heuristic algorithms can solve test problems in different sizes efficiently in a few seconds with an error of around 1%.

For future research, it's worthwhile to propose new solution approaches which can obtain the optimal solution.